\pdfoutput=1

\documentclass[11pt]{article}

\usepackage[final]{acl}

\usepackage{times}
\usepackage{latexsym}

\usepackage[T1]{fontenc}

\usepackage[utf8]{inputenc}
\usepackage{multirow}
\usepackage{microtype}

\usepackage{inconsolata}

\usepackage{graphicx}

\title{Advancing Vietnamese Information Retrieval with Learning Objective and Benchmark}

\author{Phu-Vinh Nguyen$^{1,2}$, Minh-Nam Tran$^{1,2}$, Long Nguyen$^{1,2}$\thanks{\makebox[3.1cm]{Corresponding author.}}, Dien Dinh$^{1,2}$ \\
\textsuperscript{1}Faculty of Information Technology, University of Science, Ho Chi Minh City, Vietnam \\
\textsuperscript{2}Vietnam National University, Ho Chi Minh City, Vietnam \\
\texttt{\{npvinh20,tmnam20\}@apcs.fitus.edu.vn, \{nhblong,ddien\}@fit.hcmus.edu.vn}}

\begin{document}
\maketitle
\begin{abstract}
With the rapid development of natural language processing, many language models have been invented for multiple tasks. One important task is information retrieval (IR), which requires models to retrieve relevant documents. Despite its importance in many real-life applications, especially in retrieval augmented generation (RAG) systems, this task lacks Vietnamese benchmarks. This situation causes difficulty in assessing and comparing many existing Vietnamese embedding language models on the task and slows down the advancement of Vietnamese natural language processing (NLP) research. In this work, we aim to provide the Vietnamese research community with a new benchmark for information retrieval, which mainly focuses on retrieval and reranking tasks. Furthermore, we also present a new objective function based on the InfoNCE loss function, which is used to train our Vietnamese embedding model. Our function aims to be better than the origin in information retrieval tasks. Finally, we analyze the effect of temperature, a hyper-parameter in both objective functions, on the performance of text embedding models.

\end{abstract}

\section{Introduction}
With the born of transformer architecture \cite{NIPS2017_3f5ee243} since 2017, many language models such as BERT \cite{devlin-etal-2019-bert}, GPT \cite{NEURIPS2020_1457c0d6}, and T5 \cite{t5_raffel2020exploring} have been developed and have strong performance in many natural language tasks. Furthermore, the rise of many large language models (LLMs) recently, such as Llama \cite{touvron2023llama}, Mixtral \cite{jiang2024mixtral}, Qwen \cite{qwen}, and Phi \cite{phi}, has gained strong attention for the research community due to their exceptional performance in text generation. However, LLMs have one disadvantage, they cannot access the custom data and new information to update their knowledge, which makes them unable to shift their knowledge to fit different applications. Consequently, Retrieval-Augmented Generation \cite{NEURIPS2020_6b493230} systems (or RAG) are invented to handle the problem by utilizing retrieval systems to search for relevant information from the database before feeding those information to LLMs as an extra context. This shows the necessity and importance of embedding language models for retrieval and reranking tasks in the era of LLMs.

Despite the importance of retrieval systems for LLMs, in Vietnam, the number of existing benchmarks for retrieval and reranking tasks are limited, which leads to the difficulty in comparing and assessing the performance of many Vietnamese embedding language models on those two tasks. Despite there are some Vietnamese benchmarks like ViGLUE \cite{tran-etal-2024-viglue}, ViNLI \cite{huynh-etal-2022-vinli}, VMNLU \footnote{https://github.com/ZaloAI-Jaist/VMLU.git}, and VSFC \cite{Nguyen2018UITVSFCVS}, none of them evaluate performance of language models on retrieval and reranking tasks. This paper attempts to address the need for those benchmarks by introducing a new benchmark, the Vietnamese Context Search (or the VCS) to evaluate the ability of text embedding models to search for relevant Vietnamese documents. This benchmark is constructed using existing Vietnamese datasets with modifications in their structure and tasks. Despite having a simple construction process, this benchmark effectively provides different inspections of Vietnamese text embedding models. The VCS serves as a standard and high-quality benchmark to evaluate and compare different Vietnamese embedding models on retrieval and reranking tasks.

Furthermore, this work also introduces a new training objective to train Vietnamese embedding language models on retrieval and reranking tasks. This training objective aims to yield better performance of embedding language models compared to the InfoNCE loss function, which is usually used in contrastive learning. The research will experiment with different training objectives with two training methods, including in-batch negative and curated hard-negative to compare the ability of two loss functions. Next, the evaluation of some existing Vietnamese embedding language models on the VCS benchmark is conducted to examine their ability in context search. Lastly, an empirical study is conducted to understand the effect of temperature $\tau$ in the loss function on the overall performance of embedding models. Different training methods are included in the study to further investigate the impact of temperature on the loss function.

To conclude, this work includes three primary contributions:
\begin{itemize}
    \item First, introduce a new Vietnamese benchmark, the VCS, to evaluate Vietnamese language models in their ability to search relevant documents. This benchmark evaluates models on two tasks, retrieval and reranking tasks.
    \item Second, introduce a new training objective function to train text embedding models on retrieval and reranking tasks
    \item Lastly, we conduct an empirical study to investigate the impact of temperature, a hyper-parameter, of the InfoNCE and our loss functions in the performance of embedding language models on reranking and retrieval tasks.
\end{itemize}

\section{Related Work}
In the era of large language models (LLMs), not only does the development of different generative language models such as Gemma \cite{team2024gemma}, SeaLLM \cite{Nguyen2023SeaLLMsL}, and Mamba \cite{mamba2} gains the attention from the community, but also do embedding language models, especially those support searching text documents like GTE \cite{Li2023TowardsGT}, NV-Embed \cite{Lee2024NVEmbedIT}, BGE \cite{Luo2024BGELE}, or GritLM \cite{muennighoff2024generative}, become more important due to their applications in RAG systems, which provide more context and information for LLMs to generate correct answers. Consequently, many works aim to provide a benchmark to evaluate language models on their ability in information retrieval (IR) such as BEIR \cite{thakur2021beir}, MTEB \cite{muennighoff-etal-2023-mteb}, BRIGHT \cite{BRIGHT}, and ReQA \cite{ahmad-etal-2019-reqa}. Those benchmarks advance the development of many text embedding language models and the research of natural language processing (NLP) by supporting the research community with resources to compare and evaluate text embedding models.

However, similar and comparable benchmarks for Vietnamese embedding language models are limited. While there are some benchmarks like ViGLUE~\cite{tran-etal-2024-viglue}, ViQuAD~\cite{nguyen-etal-2020-vietnamese}, ViSFD~\cite{ViSFD}, VMLU, VSMEC~\cite{VSMEC}, and VSFC~\cite{8573337}, they mostly focus on question-answering and natural language understanding aspects of language models and completely ignore the ability of language models in retrieval and reranking tasks. That leads to the difficulty in evaluating and comparing Vietnamese text embedding models in their ability of retrieve relevant information. Despite some Vietnamese embedding language models being created, without a standard benchmark on this field, the Vietnamese research community is unable to know the benefits, pros, and cons of those language models, which can lead to misleading when applying them to applications (RAG systems) or research projects.

In the early days of information retrieval, different systems were created to find relevant text information from large databases such as TF-IDF~\cite{tfidf}, BM25~\cite{bm25}, and BM25F~\cite{PrezAgera2010UsingBF}. However, those methods cannot capture the context of documents and use it for the retrieval process. Consequently, different retrieval methods using deep learning models are utilized to encode a piece of text to a vector that can present different or hidden aspects of it. Many pre-trained language models (PLMs), including BERT~\cite{devlin-etal-2019-bert}, RoBERTa~\cite{liu2019roberta}, and DeBERTa~\cite{he2020deberta}, are employed to construct new embedding models due to their capabilities in understanding natural language. Some existing embedding language models like DPR~\cite{karpukhin-etal-2020-dense} (Dense Passage Retrieval) and ColBERT~\cite{khattab2020colbert} utilize a dual encoder model structure with two separate encoders, one for queries and one for documents. Despite being fast at reference time, training two separated models would take a lot of effort and time. Meanwhile, cross-encoder models use one encoder for both queries and documents, which is more effective for the training process. Moreover, different approaches are invented to improve the performance of retrieval systems and utilize as much data as they can such as in-batch negative, which uses other examples within the same training batch as negative samples, curated hard-negative training, selecting challenging negative samples that are difficult to distinguish from positive samples, and simCSE~\cite{gao-etal-2021-simcse} pre-training method, which employs different dropout rate to create different embedding vectors of a text as positive samples.

\section{Methodology}
In this section, we explain our method to create tasks of the Vietnamese Context Search benchmark and go into detail about this benchmark. Furthermore, we introduce and explain our proposed training loss function, a modified version of the InfoNCE loss function, and our training method to create a Vietnamese text embedding model.

\subsection{Vietnamese Context Search benchmark}
Due to the lack of Vietnamese benchmarks to compare and evaluate text embedding models on information retrieval tasks, this section proposed a new Vietnamese benchmark to tackle the problem.

\subsubsection{ViMedRetrieve}
Given a database with $n$ documents $d$, the mission of a retrieval system, given a query $q$, is to retrieve documents most relevant to the query $q$ from the given database. As the number of documents $n$ increases, this task will become more challenging and require text embedding models to understand natural language to embed sequences more precisely with much information. This real-life scenario inspires us to create a new and similar benchmark to evaluate Vietnamese text embedding models.

This dataset includes $n$ different pairs of $(q,d)$, where $q$ is the question and $d$ is the document containing relevant information to answer the question $q$. In this task, the primary mission of an embedding language model, given question $q'$ as input, is to search for the expected document, which is the document $d'$ of the same pair with $q'$, after $k$ tries. This is similar to how a retrieval system would work in real-life scenarios if we consider $q$ as user input and $d$ as the document the user expects to retrieve. For further experiments on this task, we try different values of $k$ in $\{5,10,20\}$ and take accuracy when $k=5$, reflecting the ability of the embedding language model to retrieve the correct document instantly, is the primary score.

To construct this task, we re-use the ViMedAQA~\cite{tran-etal-2024-vimedaqa} dataset, a collection of Vietnamese questions and answers in healthcare, and create a new task based on it. This dataset includes four distinguished topics (drug, body part, medicine, and disease). We collect a set of questions and corresponding contexts from the dataset and use them as pairs of queries and documents for this task. To evaluate embedding models on this dataset, we use accuracy as the main metric, the model needs to search for the best $k$ documents from the whole dataset for each question, and if the model can find the relevant document within the first $k$ documents, its answer is considered to be correct and vise versa. This process creates a dataset with over 44 thousand pairs of queries and documents. The test set of this dataset, which includes over two thousand samples, is employed to evaluate embedding systems.

\subsubsection{ViRerank}
Given a query and a list of relevant and irrelevant reference texts, the target of an embedding model in the reranking task is to embed all reference texts, and then rank them based on the similarity of reference and query. This final ranking result is used to evaluate the performance of text embedding models. In this research, we utilize the mean Average Precision (mAP) metric to assess language model ability on all reranking tasks, including ViRerank.

To construct the dataset for the reranking task, we employ the ViNLI dataset, a Vietnamese benchmark for natural language inference (NLI). The ViNLI includes pairs of text pieces labeled to show their relationship, which is classified into one of four classes (entailment, contradiction, neutral, and other). The ViRerank dataset utilizes one part of each ViNLI text pair as the query and the corresponding text piece as the reference. Furthermore, each query in the ViRerank has multiple references as the ViNLI uses the same sentence for many text pairs. Positive references are chosen from text pieces labeled as entailment with the query, while negative references are taken from different labels.

The final result of this process is a new dataset with 363 samples for the test set and 367 samples for the development set, while the train set includes over 3000 samples. However, in this work, to prevent biased evaluation results toward the training and development set, we only use the test set to evaluate Vietnamese text embedding models.

\subsubsection{MNLI-R and QNLI-R}
Similar to the ViRerank dataset, we utilized two tasks from the ViGLUE dataset, MNLI and QNLI, for the reranking task. The MNLI task requires models to determine the relationship between a pair of sentences. In contrast, the QNLI task involves determining if the answer to a given question can be found in a sentence from a passage. We collect duplicated texts for each task and use them as queries just like in the ViReRank task. The \texttt{entailment} sentences (corresponding to the query) are used as positive examples and different labels are negative.

We do not employ this method for other NLI tasks of the ViGLUE dataset due to the insufficient amount of duplicated samples in those tasks. Applying this method to MNLI and QNLI creates two new sub-sets for reranking tasks, MNLI-R, with over 3.000 samples, and QNLI-R, with over 1.000 samples. Despite being the same reranking task, MNLI-R evaluates models on their ability of reranking based on context similarity while QNLI-R assesses models on their answer-searching capability.

\subsection{Training Vietnamese Embedding Model}
In this section, we introduce our training method and training objective to train a new Vietnamese embedding model for retrieval and reranking tasks.

\subsubsection{Model architecture}
Given a text sequence $x=(x_1,\dots,x_n)$ consisting of $n$ tokens, the objective is to extract information from this piece of text and map it into $R^d$, a d-dimensional space. This task can be fulfilled using an embedding model $E$ such that $e=E(x)\in R^d$ where $e$ is a presentation vector of $x$ in $R^d$.

We first use a pre-trained BERT (Bidirectional Encoder Representations from Transformers) model to extract contextual information of every token in text $x$. The output of this model is as follows:
\begin{equation}
c=LM(x)\in R^{n\times d}
\label{eqn:compute_token_context}
\end{equation}
Where the output $c$ of the language model is an embedding matrix of $n$ tokens in the sequence $x$, each token is represented by a $d$-dimentional vector.

After that, a mean pooling layer is employed to gather all contextual representations of tokens and obtain the final embedding for the entire text.
\begin{equation}
e=\frac{1}{n}\sum_1^n{c_i}
\label{eqn:compute_embedding}
\end{equation}
Where $c_i$ is the context embedding of $x_i$, the $i$-$th$ token of the sequence. This results in a $d$-dimensional vector $e$, a presentation of input $x$.

\subsubsection{Instruction training}
In retriever and re-ranking tasks, two different inputs are query and document. To handle them separately, some previous work used two embedding modules, one to encode queries and another to encode retrieved documents. This solution requires more resources during the training process as we need to train two embedding models separately.

Another solution is to apply different prompts for the query and document. By giving a hint from the input, the model can understand how to perform different calculations to compute embedding for queries and documents. This method significantly reduces the resources used to train embedding models while ensuring that the model will be trained on as much data as possible, which enhances the model's ability to comprehend the natural language. Some text embedding models such as \texttt{gte-Qwen2-7B-instruct} utilize this method and can achieve extremely high performance.

In this research, we employ instruction training to train our text embedding models. For input query, we add \texttt{<|query|>} as the prefix before feeding the whole text to the model. Meanwhile, we keep the retrieved documents the same without any modification. The difference between the two types of input lets the text embedding model know when and how to embed input query and document.

\subsubsection{Training methods}
In this work, we experiment with two different training methodologies: in-batch negatives and curated hard-negative training, and see how different training methods could affect the performance of the model on retrieval and reranking tasks.

In-batch negative sampling is a technique to improve the model's ability to differentiate the positive and negative pair of text. Given a batch of text, $x=(x_1,\dots,x_n)$ and its positive pair of text $x^+=(x^+_1,\dots,x^+_n)$, in-batch negative sampling consider all text pieces in the batch, except for the corresponding positive one, are negative. The task is to maximize the similarity of positive pairs and minimize the similarity of all the remaining negative pairs. This method has been proven to be highly resource-effective in training embedding models as it can train on $n^2$ pairs of text with a batch of $n$ pairs of text. However, as negative pairs are collected randomly during training, a negative text pair can be too obvious or not exactly negative.

Meanwhile, curated hard-negative requires the dataset to be more precise and challenging. Given a dataset item $(x,x^+,x^-)$, where $(x^+,x)$ is positive pair and the negative pair is $(x^-,x)$. Similar to in-batch negative, the target of curated hard-negative is to maximize the similarity of the positive pair and minimize those of the negative pair. The advantage of this type of training is that the negative pairs can be more challenging to differentiate, which forces the model to learn about different aspects of a text.

\subsubsection{Training objectives}
Denote $s(x,x^+)$ and $s(x,x^-)$ are predicted similarity scores of positive and negative text pairs. $p^+$ is comprehended as the probability of the positive pair. To train an embedding model on contrastive objectives and distinguish relevant documents from those that are irrelevant, one popular objective is the InfoNCE loss, which can be written as follows:
\begin{equation}
    p^+=\frac{e^{s(x,x^+)/\tau}}{e^{s(x,x^+)/\tau} + \sum{e^{s(x,x^-)/\tau}}}
\label{eqn:option_prob}
\end{equation}
\begin{equation}
L=-log(p^+)
\label{eqn:calculate_loss}
\end{equation}
The primary objective of this loss function is to increase the similarity of $(x,x^+)$ pairs while decreasing the similarity of negative text pairs. However, when the positive pair has a higher probability than other negative pairs, this training objective might still put much effort into increasing it and decreasing the likelihood of different text pairs, ignoring their relationship. With that theory, we modify the InfoNCE loss function, the idea is to lessen the loss more, which leads to slower learning speed, as $p^+$ gets larger. This target can be easily fulfilled by multiplying the final InfoNCE loss with $(1-p^+)$, resulting in the loss function in Equation~\ref{eqn:our_loss}:
\begin{equation}
L_{ours}=-log(p^+) (1-p^+)
\label{eqn:our_loss}
\end{equation}
The term $(1-p^+)$ added in the function is used as an extra weight to the loss function, which gets smaller as the probability of the correct pair is higher, slowing down the learning speed of the model on correct examples. This extra weight prevents the over-learned scenario of the original loss function by reducing the gradient from the loss value to the model's weights on those samples.

\subsubsection{Training datasets}
We create two different training datasets for two different training methods, training with in-batch negative and training with curated hard-negative. Despite having different structures, those datasets share the same data-collecting method. We first collect data from three primary resources, the Vietnamese NewsSapo dataset~\cite{duc2024towards}, the Binhvq News Corpus\footnote{https://github.com/binhvq/news-corpus}, and the Vietnamese version of the QQP triplet~\cite{viqqp_triplet}. The dataset is summarized in Table~\ref{tab:train_dataset}.

\begin{table}[htpb]
  \centering
  \begin{tabular}{lc}
    \hline
    \textbf{Dataset} & \textbf{Number of samples} \\
    \hline
    BKAINewsCorpus     & 1.5M\\
    Vietnamese QQP triplet     & 101K\\
    Binhvq News Corpus     & 1M\\\hline
  \end{tabular}
  \caption{Dataset summarization for the training set before filtering samples based on text length}
  \label{tab:train_dataset}
\end{table}
Next, we filter this dataset based on text length. Then, to prepare a dataset for in-batch negative training, we remove all negative examples from each data sample, leaving only a text pair of anchor and positive text. Meanwhile, for curated hard-negative training, we keep the original negative examples while adding negative samples to those text groups that do not have any. We do this by randomly selecting a text piece in the dataset that does not belong to the original group. Although this method may not provide a difficult and high-quality training curated hard-negative dataset, the text embedding models can still learn the relationship between the positive and negative text pairs.

\section{Experiments}
In this section, we compare our modified loss function with the InfoNCE loss function with different training methods. Furthermore, we also evaluate the performance of our embedding language models from the previous step and compare them with some existing Vietnamese embedding models on retrieval and reranking tasks of our benchmark. Lastly, we investigate the effect of temperature $\tau$ on the performance of embedding models as they are trained with ours and the InfoNCE loss function.

\subsection{Comparision of training objectives and training method}
In this experiment, we fine-tune the pre-trained BERT-based embedding model \footnote{https://huggingface.co/sentence-transformers/all-MiniLM-L6-v2} on the Vietnamese dataset. Despite this model being pre-trained on English datasets, its performance on our Vietnamese benchmark is reasonably high. Furthermore, its small size can provide an empirical study and comparison of different training methods without requiring much computational resources.

We train text embedding models using two loss functions, ours and the InfoNCE loss function. Furthermore, two training methods, including in-batch negative and curated hard-negative training, are employed in this experiment. Finally, we evaluate those models on our benchmark. The result of this experiment is summarized in Table~\ref{tab:experiment_rerank} and Table~\ref{tab:experiment_retrieve}.

\begin{table}[htpb]
\centering
\begin{tabular}{ll| l l l}
\hline
\multicolumn{2}{l}{}& \textbf{ViNLI} & \textbf{MNLI-R} & \textbf{QNLI-R} \\ \hline
\multicolumn{2}{l|}{baseline} & 62.42 & 78.92 & \textbf{87.06} \\ \hline
\multicolumn{1}{l|}{\multirow{2}{*}{InfoNCE}} & IB & 62.07 & 75.61 & 85.26 \\
\multicolumn{1}{l|}{} & HN & 66.27 & 83.86 & 85.56 \\ \hline
\multicolumn{1}{l|}{\multirow{2}{*}{ours}} & IB & 63.24 & 77.15 & 86.22 \\
\multicolumn{1}{l|}{} & HN & \textbf{67.86} & \textbf{84.51} & 86.04 \\ \hline
\end{tabular}
\caption{Experiment results on reranking tasks using the mAP score. \textbf{IB} denotes the in-batch negative training method, and \textbf{HN} refers to curated hard-negative training. Results are presented as percentages.}
\label{tab:experiment_rerank}
\end{table}

From the experiment results of Table~\ref{tab:experiment_rerank}, training with hard-negative examples results in better performance compared to the in-batch negative training method for all reranking tasks. Furthermore, in some reranking tasks, training with the in-batch negative method might degrade the performance of the model on this task. Next, our training objective reproduces better performance in all reranking tasks and all methods compared to the InfoNCE loss function despite there is still a degradation in task QNLI-R as we compare with the baseline model. Lastly, the baseline model, despite only being trained on the English datasets, has a relatively high performance. As MNLI and QNLI tasks in the ViGLUE dataset are translations from the GLUE benchmark, some English structural patterns, and similar terminology may be retained in the translated versions, which could explain why the baseline model performs well on these tasks despite having limited knowledge of Vietnamese.

\begin{table}[htpb]
\centering
\begin{tabular}{ll|lll}
\hline
\multicolumn{2}{l|}{\multirow{2}{*}{}} & \multicolumn{3}{c}{\textbf{ViMedRetrieve}} \\ \cline{3-5} 
\multicolumn{2}{l|}{} & \multicolumn{1}{l}{k@5} & \multicolumn{1}{l}{k@10} & k@20 \\ \hline
\multicolumn{2}{l|}{baseline} & \multicolumn{1}{l}{0.20} & \multicolumn{1}{l}{0.37} & 0.53 \\ \hline
\multicolumn{1}{l|}{\multirow{2}{*}{InfoNCE}} & IB & \multicolumn{1}{l}{0.25} & \multicolumn{1}{l}{0.32} & 0.36 \\
\multicolumn{1}{l|}{} & HN & \multicolumn{1}{l}{0.24} & \multicolumn{1}{l}{0.27} & 0.29 \\ \hline
\multicolumn{1}{l|}{\multirow{2}{*}{ours}} & IB & \multicolumn{1}{l}{0.26} & \multicolumn{1}{l}{\textbf{0.46}} & \textbf{0.59} \\
\multicolumn{1}{l|}{} & HN & \multicolumn{1}{l}{\textbf{0.30}} & \multicolumn{1}{l}{0.44} & 0.50 \\ \hline
\end{tabular}
\caption{Experiment results on retrieval tasks with varying numbers of retrieved items. \textbf{IB} is in-batch negatives, and \textbf{HN} refers to curated hard negatives. Results are presented as percentages based on the accuracy metric.}
\label{tab:experiment_retrieve}
\end{table}

However, the result from Table~\ref{tab:experiment_retrieve} shows the opposite: for both objective functions, the performance of the in-batch negative method is higher than that of the hard-negative training method. With our training objective applied, the in-batch negative training method can raise a better result with $k=10$ and $k=20$ while unable to surpass when $k=5$, this shows that the in-batch negative method has better performance if we want to find correct documents with a large number of finding at a time. Moreover, the result of our objective function is still higher than that of the infoNCE loss function with multiple values of $k$ and different training methods. Furthermore, the low results of other methods on this task depict the difficulty of this task. Lastly, using the infoNCE loss function degrades significantly the performance of the baseline model in the retrieval task, this can be a consequence of low-quality training data in the case of curated hard-negative training. However, in in-batch negative training, the employed loss function plays a crucial role in this reduced performance.

\subsection{Comparision of Vietnamese embedding models}
\begin{table}[htpb]
    \centering
    \begin{tabular}{lc}
    \hline
        \textbf{Model} & \textbf{Parameters}  \\
    \hline
         SimCSE & 130M \\
         Bi-encoder & 130M \\
         Sbert & 130M \\
         ours & 20M \\
    \hline
    \end{tabular}
    \caption{Number of parameters of Vietnamese embedding language models in the experiment}
    \label{tab:model_size}
\end{table}

This experiment will explore the ability of Vietnamese embedding language models to retrieve and rerank tasks by evaluating the VCS benchmark. The models used in this experiment includes \texttt{sup-SimCSE-VietNamese-phobert-base}\footnote{https://huggingface.co/VoVanPhuc/sup-SimCSE-VietNamese-phobert-base}, \texttt{vietnamese-bi-encoder}\footnote{https://huggingface.co/bkai-foundation-models/vietnamese-bi-encoder}, \texttt{vietnamese-sbert}\footnote{https://huggingface.co/keepitreal/vietnamese-sbert}, and our models. It is worth noticing that the three first models in the list are trained based on \texttt{phoBERT} with 135M parameters and our experimental model has just 20M, this is stated in Table~\ref{tab:model_size}. The result of this experiment is reported in Table~\ref{tab:vmodel_rerank} and Table~\ref{tab:vmodel_retrieve}.

\begin{table}[htpb]
\begin{tabular}{l|lll}
\hline
                 & \textbf{ViRerank} & \textbf{MNLI-R} & \textbf{QNLI-R} \\ \hline
SimCSE           & \textbf{69.46} & \textbf{87.74} & 88.50          \\
Bi-encoder       & 65.41 & 82.10 & \textbf{90.30} \\
Sbert            & 66.9 & 83.57 & \underline{88.79} \\
ours             & \underline{67.86} & \underline{84.51} & 86.04          \\ \hline
\end{tabular}
\caption{Vietnamese embedding models comparison on reranking tasks, measured by mAP metric. \textbf{Bold} text expresses the highest score, \underline{Underline} highlight the second highest score.}
\label{tab:vmodel_rerank}
\end{table}

From the evaluation result in Table~\ref{tab:vmodel_rerank}, model \texttt{sup-SimCSE-VietNamese-phobert-base} achieves the highest score on the ViRerank and MNLI-R tasks with a score of 69.46 and 87.74 respectively. Our model comes in second place in the same tasks with 67.86 on ViRerank and 84.51 on MNLI-R. For the last reranking task, QNLI-R, model \texttt{vietnamese-bi-encoder} has the highest score with 88.79 while model \texttt{vietnamese-sbert} is in the second place with 88.79. 

\begin{table}[htpb]
    \centering
\begin{tabular}{l|lll}
\hline
                 & \multicolumn{3}{c}{\textbf{ViMedRetrieve}} \\ \cline{2-4}
                 & \multicolumn{1}{l}{k@5} & \multicolumn{1}{l}{k@10} & k@20 \\ \hline
SimCSE           & \multicolumn{1}{l}{0.09} & \multicolumn{1}{l}{0.11} & 0.12 \\ 
Bi-encoder       & \multicolumn{1}{l}{\underline{0.25}} & \multicolumn{1}{l}{\textbf{0.45}}          & \textbf{0.73} \\
Sbert & \multicolumn{1}{l}{0.18}          & \multicolumn{1}{l}{0.26}          & 0.32          \\
ours             & \multicolumn{1}{l}{\textbf{0.30}} & \multicolumn{1}{l}{\underline{0.44}} & \underline{0.50}          \\ \hline
\end{tabular}
\caption{Vietnamese embedding models comparison on retrieval task, measure by accuracy. \textbf{Bold} text expresses the highest score, \underline{Underline} highlight the second highest score.}
\label{tab:vmodel_retrieve}
\end{table}

Despite having high performance on reranking tasks, the performance on the retrieval task of \texttt{sup-SimCSE-VietNamese-phobert-base} model is significantly lower compared to other Vietnamese embedding models. Meanwhile, \texttt{vietnamese-bi-encoder} can achieve the highest score when the number of retrieved items $k$ is set to 10 or 20, and is the second highest when $k=5$. Our model, on the other hand, gets the highest score as $k=5$ and comes in second place as the number of retrieved items $k$ increases to 10 and 20.

From the results on retrieval and reranking tasks, \texttt{sup-SimCSE-VietNamese-phobert-base} presents a strong ability in reranking tasks, which contain a small number of text. However, in retrieval tasks with a large amount of text, \texttt{vietnamese-bi-encoder} tend to have better performance than different embedding models. Furthermore, our model, with just over 20 million parameters, is on par with three existing Vietnamese embedding language models with larger sizes.

\subsection{Affect of temperature on performance}
\begin{figure*}[htpb]
    \centering
    \includegraphics[width=0.95\textwidth]{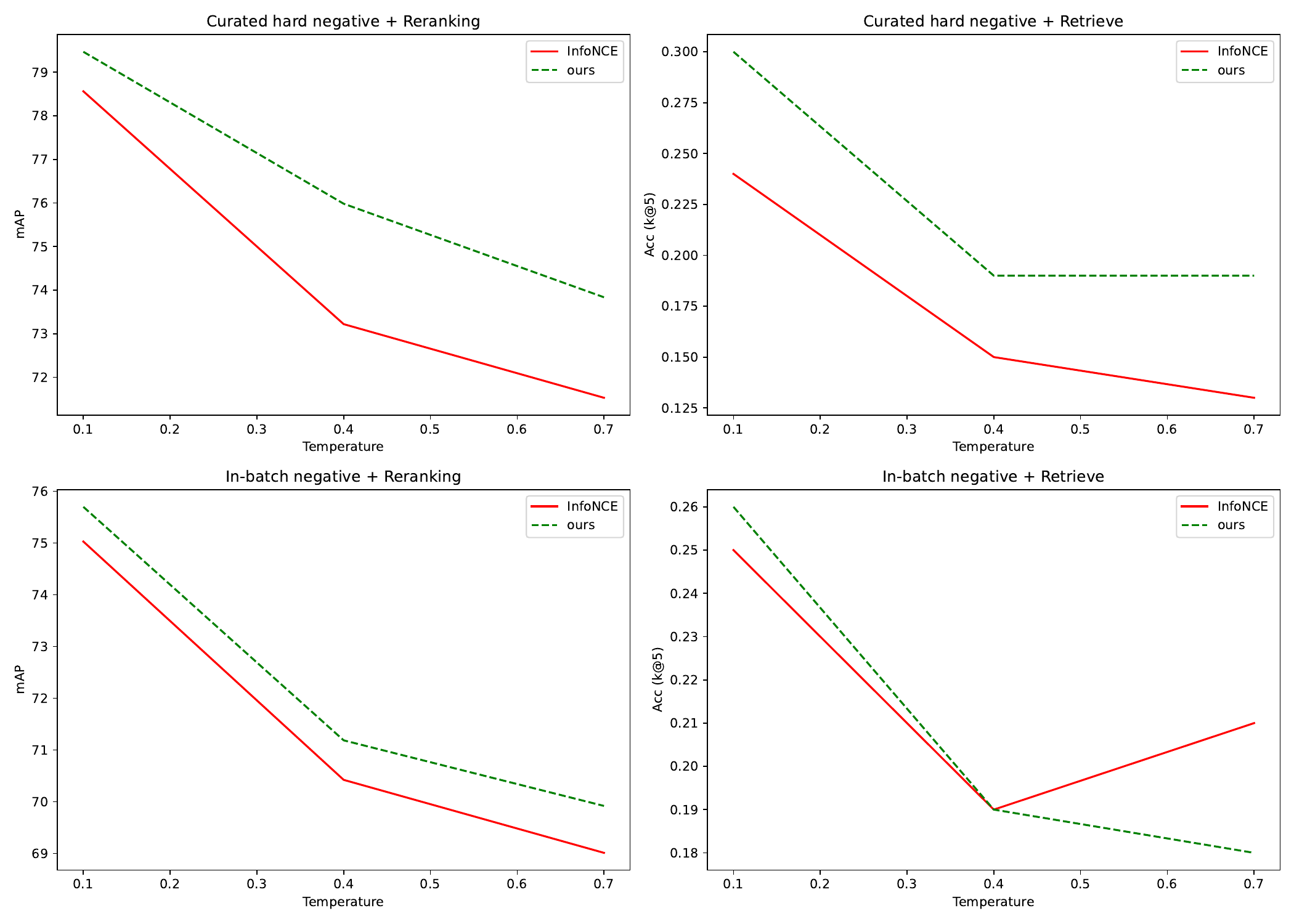}
    \caption{The impact of temperature $\tau$ on the model's performance on two tasks: retrieval and reranking, along with two training methods: in-batch negative and curated hard-negative.}
    \label{fig:temperature}
\end{figure*}

This experiment explores the different values of temperate ($\tau={0.1,0.4,0.7}$) in the InfoNCE loss function and our loss function. The result of this experiment is visualized in Figure~\ref{fig:temperature}.

From Figure~\ref{fig:temperature}, the performance of embedding models on reranking tasks decreases as the temperature $\tau$ increases. This phenomenon happens for both training objectives (InfoNCE loss and our loss function) as well as for both training methods (curated hard-negative and in-batch negative). Furthermore, the performance of models on retrieval tasks significantly decreases when the temperature increases from $0.1$ to $0.4$. However, when the temperature increases from $0.4$ to $0.7$, different behaviors are recorded for different combinations of training objectives and training methods. For models trained on curated hard-negative with the InfoNCE loss function and models trained on in-batch negative with our loss function, their performance on retrieval tasks slightly decreased. Meanwhile, the performance of the model trained on the in-batch negative with the InfoNCE loss will increase. Finally, the model trained on curated hard-negative with our loss function remains consistent performance when temperature increases from $0.4$ to $0.7$.

It is also important to notice that our loss function raises better performance on both retrieval and reranking tasks with different temperatures, except for the retrieval task with in-batch negative training when the InfoNCE loss has better performance with $\tau=0.7$. Furthermore, this experiment shows that the temperature should be low for text embedding models to perform well on retrieval and reranking.

\section{Conclusion}
This work constructs the Vietnamese Context Search benchmark to evaluate Vietnamese embedding language models on retrieval and reranking tasks, with three validation datasets (ViMedRetrieve, ViRerank, and ViGLUE-R). Moreover, this work presents a new training objective function, which performs better than the InfoNCE loss function in reranking and retrieval tasks. Lastly, we evaluate the performance of some Vietnamese embedding language models on our benchmark and experiment to study the effect of temperature $\tau$ on the performance of embedding models with different training methods.

\section{Limitation and Future works}
One limitation of this work is the difficulty of the ViMedRetrieve dataset, which makes the results of many Vietnamese embedding language models extremely low. Moreover, the evaluation score of ViMedRetrieve is conducted based on the accuracy of different numbers of retrieved documents. Despite providing more detail about the model's performance, this metric poorly summarizes the model's overall performance on the retrieval task. Future works aim to add a new metric to evaluate the overall model's performance on this task.


\bibliography{custom}

\appendix

\section{Appendix}
\label{sec:appendix}

\subsection{Hardware Resources}
This research uses the free NVIDIA Tesla P100 PCIe 16 GB 824 provided by Kaggle.

\subsection{Hyperparameters}
The hyper-parameters used in the training process are reported in Table~\ref{tab:hyperparam}.
\begin{table}[htpb]
    \centering
    \begin{tabular}{c|c}
    \hline
         \textbf{Hyper-parameter} & \textbf{Value}  \\ \hline
         Batch size & 32 \\
         Learning rate & 5e-5 \\
         Max sequence length & 224 \\
         Epochs & 3 \\
         Temperature & \{0.1, 0.4, 0.7\} \\ \hline
    \end{tabular}
    \caption{Hyper-parameters used in in-batch negative and curated hard-negative training}
    \label{tab:hyperparam}
\end{table}
\subsection{Running Time}
The running time for the in-batch negative training is 4 hours and 45 minutes while training with the curated hard-negative training method requires 7 hours and 30 minutes.

\subsection{Datasets and models}
The datasets and models used in this paper are publicly available on Hugging Face\footnote{\url{https://huggingface.co/ContextSearchLM}} and GitHub\footnote{\url{https://github.com/phuvinhnguyen/VietnameseTextSearch}}.

\end{document}